\documentclass[letterpaper]{article} 
\usepackage{aaai2026}                
\usepackage{times}                   
\usepackage{helvet}                  
\usepackage{courier}                 
\usepackage[hyphens]{url}            
\usepackage{graphicx}                
\usepackage{natbib}                  
\usepackage{caption}                 

\usepackage{comment}
\usepackage{spverbatim}
\usepackage{listings}
\usepackage{mdframed}
\usepackage{fancyvrb}
\usepackage{multirow}
\usepackage{multicol} 
\usepackage{tcolorbox}
\usepackage{colortbl}  
\usepackage{pifont}
\usepackage{makecell}
\usepackage{graphicx}
\usepackage{caption}
\usepackage{subcaption}
\usepackage{booktabs}
\usepackage{alltt}
\usepackage{rotating}
\usepackage{array}
\usepackage{multirow}
\usepackage{amsfonts}
\usepackage{amsmath}
\usepackage{enumitem}

\definecolor{vlgray}{rgb}{0.95,0.95,0.95}
\usepackage[T1]{fontenc}

\usepackage{pifont}
\newcommand{\emptycirc}{\ding{109}}
\newcommand{\halfcirc}{\ding{119}}
\newcommand{\fullcirc}{\ding{108}}

\usepackage{url}

\usepackage{xcolor}
\definecolor{niravblue}{RGB}{0,119,190}
\definecolor{navyblue}{rgb}{0.0,0.0,0.5}

\newcommand{\comments}{}
\ifdefined\comments
  \newcommand{\gang}[1]{\textcolor{blue}{\textbf{Gang:}~#1}}
  \newcommand{\nirav}[1]{\textcolor{niravblue}{\textbf{Nirav:}~#1}}
  \newcommand{\arth}[1]{\textcolor{green}{\textbf{Arth:}~#1}}
  \newcommand{\fixed}[1]{\textcolor{red}{#1}}
\else
  \newcommand{\gang}[1]{}
  \newcommand{\nirav}[1]{}
  \newcommand{\arth}[1]{}
  \newcommand{\fixed}[1]{}
\fi

\usepackage{comment}
\newcommand{\para}[1]{\vspace{0.7em}\noindent\textbf{#1}\ }

\usepackage{booktabs}
\usepackage{multirow}

\title{Beyond BeautifulSoup: \\
Benchmarking LLM-Powered Web Scraping for Everyday Users}

\author{
Arth Bhardwaj\textsuperscript{\rm 1},
Nirav Diwan\textsuperscript{\rm 2},
Gang Wang\textsuperscript{\rm 2}
}
\affiliations{
\textsuperscript{\rm 1}Saint Francis High School, Mountain View, CA\\
\textsuperscript{\rm 2}University of Illinois Urbana--Champaign\\
arthbhardwaj1234@gmail.com, ndiwan2@illinois.edu, gangw@illinois.edu
}

\begin{document}
\maketitle

\pagestyle{plain}
\thispagestyle{plain}

\begin{abstract}
Web scraping has historically required technical expertise in HTML parsing, session management, and authentication circumvention, which limited large-scale data extraction to skilled developers. We argue that large language models (LLMs) have democratized web scraping, enabling low-skill users to execute sophisticated operations through simple natural language prompts. While extensive benchmarks evaluate these tools under optimal expert conditions, we show that without extensive manual effort, current LLM-based workflows allow novice users to scrape complex websites that would otherwise be inaccessible. We systematically benchmark what everyday users can do with \textit{off-the-shelf} LLM tools across 35 sites spanning five security tiers, including authentication, anti-bot, and CAPTCHA controls. We devise and evaluate two distinct workflows: (a) \textit{LLM-assisted scripting}, where users prompt LLMs to generate traditional scraping code but maintain manual execution control, and (b) \textit{end-to-end LLM agents}, which autonomously navigate and extract data through integrated tool use. Our results demonstrate that end-to-end agents have made complex scraping accessible—requiring as little as a single prompt with minimal refinement ($<5$ changes) to complete workflows. We also highlight scenarios where LLM-assisted scripting may be simpler and faster for static sites. In light of these findings, we provide simple procedures for novices to use these workflows and gauge what adversaries could achieve using these.
\end{abstract}


\section{Introduction}
\label{sec:intro}
The democratization of web scraping capabilities represents a fundamental shift in who can extract data from the web at scale. Historically, effective data extraction required specialized knowledge of HTML parsing, request handling, anti-bot circumvention, and complex debugging workflows~\cite{dikilitacs2023performance,stafeev2024sok,zeber2020representativeness,nikiforakis2021goodbot,azad2020webrunner}. 
This technical barrier served as a natural filter, limiting large-scale scraping operations to developers with substantial expertise and resources.

The emergence of LLMs and consumer-facing agentic frameworks has dramatically lowered these barriers. Today, users with only rudimentary technical skills can potentially execute sophisticated scraping operations using \textit{off-the-shelf} tools. LLMs can now generate scraping code, debug failures, and even provide guidance on bypassing common protections. Agents can provide end-to-end automation by navigating websites, filling forms, and adapting to dynamic content without requiring users to understand the underlying technical complexity, thereby scaling both benign automation and potential abuse against anti-bot and access-control mechanisms~\cite{agentbench2023, xie2024osworld}.

\begin{figure*}[t]
    \centering
\includegraphics[width=\textwidth]{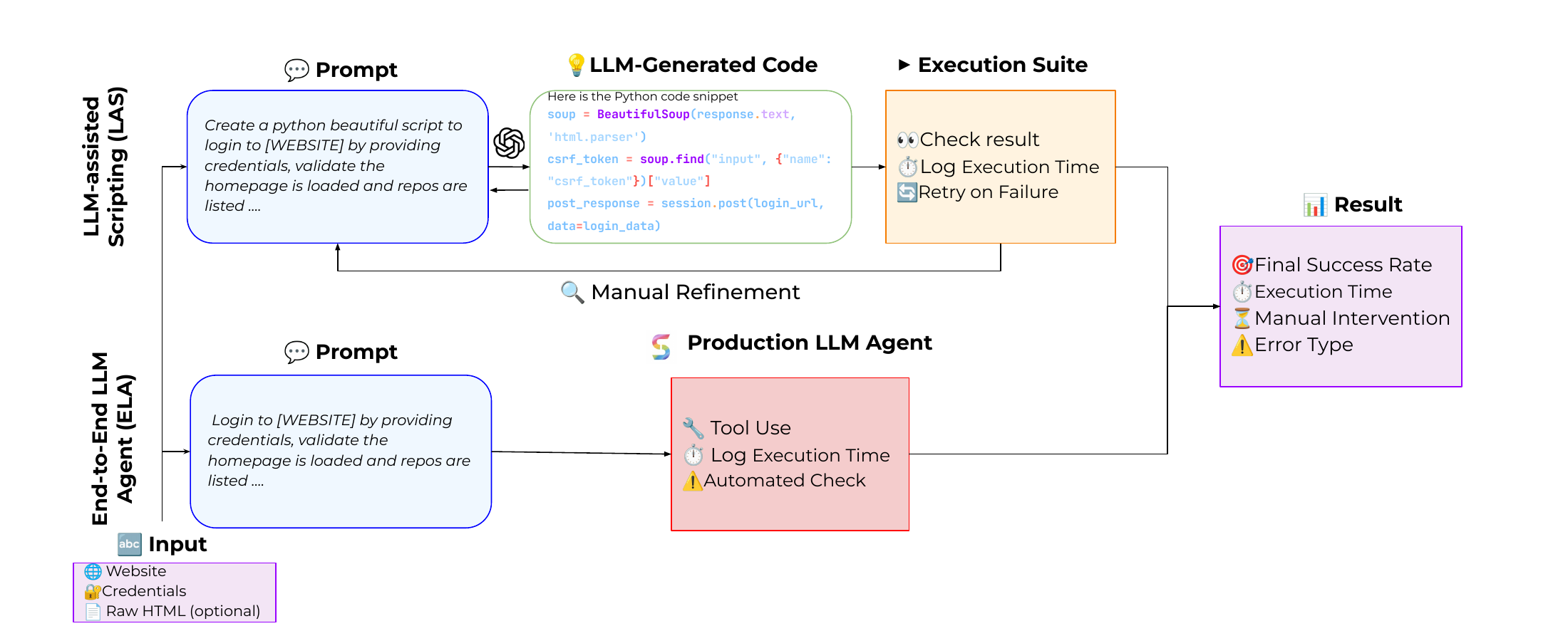} \caption{\textbf{Benchmark for non-expert web scraping}. We introduce a benchmark that evaluates what non-expert users can achieve with off-the-shelf tools, modeling two workflows: (i) \textit{LLM-assisted Scripting (LAS)} and (ii) \textit{End-to-End LLM Agent (ELA)}. LAS: the LLM drafts code that the user executes and manually refines; ELA: a production agent plans and acts via internal tools.}
    \label{fig:workflow}
\end{figure*}

Despite the widespread availability of these tools, we lack an understanding of the democratization made possible by them. Existing benchmarks focus heavily on best practices and optimal configurations~\cite{agentbench2023, xie2024osworld}, evaluating what these tools can accomplish under ideal conditions with expert guidance. However, there is an existing gap between these optimized evaluations and the reality of how non-expert users actually deploy scraping tools in practice, with minimal customization, limited debugging skills, and reliance on default settings.
Addressing this gap is critical for estimating abuse feasibility and for adopting practical defensive measures. 
While traditional scraping frameworks remain the backbone of many data extraction operations, their complexity may now be unnecessary, given the emergence of LLM Agent alternatives~\cite{bohra2025weblists,huang2024autoscraper,xie2024osworld}.
Conversely, end-to-end LLM agents, while more intuitive, may suffer from reliability and performance issues~\cite{agentbench2023,zhouwebarena}.

\para{Motivation.} The motivation for this study is to empirically assess this democratization by measuring what novice users can actually accomplish using \textit{off-the-shelf} tools on webpages with varying technical security complexities. 
By comparing traditional and agentic approaches under realistic constraints (limited time, budget, and expertise), we aim to establish whether modern web scraping has indeed become accessible for low-skill scenarios, including on security-sensitive sites that deploy authentication, anti-bot checks, and CAPTCHA.
These workflows yield an empirical estimate of what novice adversaries can achieve, and in turn inform the practices defenders should prioritize and adopt.

\para{Our Approach.} To measure this, we developed a benchmarking framework to evaluate both traditional scraping tools and emerging LLM agents under realistic web interaction scenarios. The benchmark compares two primary workflows {Figure~\ref{fig:workflow}}: (a) the \textit{LLM-assisted Scripting}, where users prompt LLMs to generate Python scripts that apply traditional scraping libraries BeautifulSoup and Scrapy but execution and configuration remain manual; (b) the \textit{End-to-End LLM Agent}, where production LLM agents (such as Claude and Simular.ai) are prompted to complete login and scraping tasks directly, with their own embedded tool use, retries, and error detection.

Prior work~\cite{lotfi2021web, huang2025prites, brach2025ghosts} primarily focuses on best practices for the web scraping workflows. 
We instead measure two things: (1) traditional \textit{correctness} metrics, and
(2) \textit{ease of use} metrics which include the execution time and manual effort needed to complete the scraping task.
To stress-test across diverse scenarios, we also select 35 websites across five progressively challenging categories of websites, ranging from static HTML websites to categories that are traditionally hard to automate for novice users (e.g., CAPTCHA).

Our results show that new workflows lower the barrier to web scraping for novice users, with even a single prompt often sufficient to handle sites that involve complex authentication or dynamic content. 
The effectiveness of these approaches, however, depends critically on site complexity. Agents are particularly effective at scraping from websites that traditional scripting libraries fail to support. However, for static websites, agents are overkill.

\para{Contributions.} We make the following key contributions:  

\begin{itemize}

\item \textbf{Benchmark Capturing Novice Use.} We introduce a benchmark spanning five progressively challenging web-interaction categories that captures off-the-shelf use of LLMs for scraping.

\item \textbf{Workflow Protocols and Unified Metrics.} We formalize two contrasting workflows for web scraping: LLM-assisted Scripting (LAS) and End-to-End LLM Agent (ELA), and define comparable metrics measuring both success and ease-of-use.

\item \textbf{Empirical Findings and Practical Guidance.} We conduct a constrained evaluation (time, success rate, and effort taken) that quantifies the accessibility–reliability trade-off between LAS and ELA across difficulty tiers, and we provide actionable recommendations on when to prefer each workflow.

\item \textbf{Open-Source.} We open-source our code, prompts and generated scripts for the community to use and contribute. Our code is available here\footnote{https://github.com/arthbhardwaj04/LLMPoweredWebScraping}. 

\end{itemize}

\section{Background and Related Work}
\label{sec:background}

Web scraping has evolved significantly over the past two decades, transitioning from simple HTML parsing on static websites to complex automation on highly interactive, JavaScript-driven applications. This evolution has been driven by changes in how web applications deliver content and protect resources~\cite{nikiforakis2021goodbot,azad2020webrunner,stafeev2024sok,zeber2020representativeness}. Modern sites frequently employ asynchronous data loading, client-side rendering, and increasingly sophisticated security mechanisms to deter automated access. Two primary categories of scraping approaches have emerged in response to these challenges: \textit{LLM-assisted Scripting} and \textit{End-to-End LLM Agent}. This section surveys the capabilities and limitations of each category, as well as relevant research on CAPTCHA-solving and benchmarking.

\para{Traditional Web Scraping Tools.} Traditional scraping frameworks such as {\em BeautifulSoup} and {\em Scrapy} operate by directly parsing the HTML of target web pages~\cite{richardson2024beautifulsoup,scrapydocs2024}. 
\textit{BeautifulSoup} is a lightweight Python library optimized for quick HTML/XML parsing and extraction using CSS selectors.
\textit{Scrapy} is a more comprehensive framework supporting asynchronous requests, structured crawling rules, and native export to CSV/JSON formats. 

These tools excel when the page structure is stable and predictable, the content is delivered primarily in static HTML, and large-scale batch scraping is required with minimal runtime overhead~\cite{dikilitacs2023performance,pochat2018tranco,cuevas2022marketplace}. However, they encounter limitations on modern, interactive websites that require: (1) client-side rendering via JavaScript, (2) multi-step user authentication flows, and (3) dynamic CAPTCHA or behavioral verification.
While headless browsers (e.g., Selenium, Playwright) can extend the reach of traditional tools~\cite{selenium2023,microsoft2024playwright}, integration complexity and execution time typically increase.

\para{LLM Web Agents.} Recent advances in \textit{Large Language Models (LLMs)} have enabled the creation of autonomous web agents that interact with web applications in a human-like fashion. Tools such as {\em Claude} and {\em Simular.ai}~\cite{anthropic2025claude,agashe2024agents,xie2024osworld} can: (1) visually interpret and reason about page layouts, (2) navigate multi-step workflows, (3) click buttons, fill forms, and follow contextual instructions, and (4) solve certain types of CAPTCHA.

Unlike traditional DOM-parsing, which depends on fixed structural selectors, LLM agents adapt to dynamic and less predictable environments. They can operate without explicit code for each page element, instead relying on natural-language instructions and visual analysis~\cite{yehudai2025survey,agentbench2023}. However, these benefits come with trade-offs such as potentially higher execution time and variability in output due to non-deterministic reasoning.

\para{CAPTCHA and Anti-Bot Mechanisms.} Many websites deploy \textit{Completely Automated Public Turing tests to tell Computers and Humans Apart} (CAPTCHAs) and other anti-bot techniques to safeguard access~\cite{vonahn2003captcha}. Common mechanisms include static image-based CAPTCHAs (e.g., character recognition), interactive puzzles (e.g., image selection, slider movement), behavioral verification (e.g., mouse movement, scroll patterns), and device/browser fingerprinting and rate limiting.

Traditional scraping tools often encounter difficulties bypassing these protections without the aid of third-party services. Some LLM agents, particularly those capable of visual analysis, can solve simpler challenges, but their success rate decreases significantly with more complex or obfuscated CAPTCHA designs~\cite{deng2024oedipus}. Recently, researchers also explored using LLMs to solve CAPTCHAs, which, however, still suffer from long resolving delays~\cite{teoh2025arecapchas}.

\para{Related Benchmarking Efforts.}
Prior benchmarking studies~\cite{ agentbench2023, xie2024osworld} have compared scraping tools based on execution speed, data accuracy, and ease of integration. However, most have focused solely on traditional frameworks and have not examined modern LLM-powered agents in realistic, security-intensive environments for everyday users. While a few studies~\cite{luo2025open,ding2025illusioncaptcha} have included sites requiring OAuth/MFA authentication or dynamic CAPTCHA solving in their evaluation set, they do not mention the level of expertise of the user.

Our work addresses these gaps by conducting a systematic, side-by-side benchmark of {\em traditional} and {\em LLM-based approaches} across 35 websites spanning five representative categories. This evaluation not only measures \textit{success rate} and \textit{execution time} but also considers \textit{setup complexity}, \textit{output accuracy}, and \textit{CAPTCHA-handling capability}, providing actionable guidance for practitioners selecting scraping solutions in modern web environments.

\section{Threat Model}
\label{sec:threat-model}

Our threat model assesses whether actors with rudimentary technical knowledge can use crawling tools and LLMs in an \textit{off-the-shelf} manner to execute web scraping operations. We intentionally focus on low-skill scenarios to establish a conservative baseline: if users with limited expertise can perform data extraction using readily available tools, this demonstrates that scraping capabilities have become sufficiently {\em democratized}.

\para{Assumptions.} We assume the actor possesses basic skills, including running Python scripts, using LLMs, and working with consumer-facing agentic frameworks. Additionally, the actor can create login accounts and manage credentials. However, the actor lacks deep expertise in scraping library nuances and does not employ advanced prompt engineering best practices. Notably, even for writing the traditional scraping workflows, the actor uses the assistance of LLMs. The actor also operates under resource constraints---limited time, budget, and computational power---and therefore relies on straightforward, accessible tools rather than developing sophisticated solutions.

\para{Implications.}
The threat lies in how easily scraping workflows (prompt $\rightarrow$ code $\rightarrow$ run $\rightarrow$ extract) can be replicated with limited oversight. 
Thus, through these tools, scraping becomes a capability amplifier: enabling both defenders (e.g., researchers detecting fraud) and attackers (e.g., spammers bypassing controls).

\section{Methodology}
\label{sec:methodology}

Our objective was to develop a technically diverse benchmarking pipeline that could evaluate tool performance across increasingly complex web scraping challenges.

\subsection{Website Selection Criteria}
\label{subsec:site-selection}

We selected 35 websites that employ protection mechanisms commonly encountered in real-world scraping scenarios, representing realistic targets for actors with basic technical skills and limited scraping experience. The selection was grounded in three key criteria:

\begin{itemize}
    \item \textbf{Realism:} All sites represent common scraping targets in research, journalism, e-commerce, or data science.
    \item \textbf{Diversity of Authentication:} We include basic login forms, token-based authentication, MFA/OAuth-protected flows, and anti-bot protections.
    \item \textbf{Technical Complexity:} Selected sites use varied frontend stacks (e.g., React, Angular, server-side rendering) and include protections like Cloudflare, rate-limiting, or CAPTCHA.
\end{itemize}

To evaluate how these actors would perform against increasing levels of defense, we classified these websites into five distinct scraping difficulty tiers mentioned in Table \ref{tab:web-scraping-comparison}:

\begin{table*}[htbp]
\centering
\small
\begin{tabular}{@{}p{2.3cm}>{\centering\arraybackslash}p{2cm}>{\centering\arraybackslash}p{2.8cm}p{8cm}@{}}
\toprule
\textbf{Web Page Type} & \textbf{Technical Complexity} & \textbf{Authentication \& Protection Requirements} & \textbf{Websites} \\
\midrule

Simple HTML & \emptycirc & \emptycirc & 
The Hacker News, 
Wikipedia, 
Investopedia, 
Goodreads \\
\midrule

Complex HTML & \halfcirc & \emptycirc & 
CNN, 
BBC, 
New York Times, 
Reuters, 
National Geographic, 
NPR, 
Vimeo \\
\midrule

Simple Authentication & \halfcirc & \halfcirc & 
PayPal, 
eBay, 
edX, 
Coursera, 
Figma, 
YouTube, 
Netflix, 
Apple, 
Spotify, 
GitHub \\
\midrule

Complex Authentication & \halfcirc & \fullcirc & 
Facebook, 
Reddit, 
Microsoft, 
Ticketmaster, 
Booking.com, 
Macy’s, 
Bitly,
Amazon, 
Slack, 
Expedia
\\

\midrule

CAPTCHA & \fullcirc & \fullcirc & 
reCAPTCHA v2 Checkbox Demo, 
2Captcha Normal Demo, 
reCAPTCHA v3 Scores Demo, 
2Captcha v3 Enterprise Demo, 
Geetest Adaptive CAPTCHA Demo \\

\bottomrule
\end{tabular}
\caption{Web scraping complexity comparison across dimensions of technical complexity and authentication/protection requirements, with representative examples. Circle symbols indicate intensity levels: empty (\emptycirc) = low/none, half-filled (\halfcirc) = medium, filled (\fullcirc) = high.}
\label{tab:web-scraping-comparison}
\end{table*}

\begin{enumerate}
\item \textbf{Simple HTML}: Public-facing pages with simple, static HTML structures. No login, JavaScript execution, or client-side rendering is required.
\item \textbf{Complex HTML}: Static pages that are deeper in structure, featuring nested tables, multi-column layouts, or dynamic class names.
\item \textbf{Simple Authentication}: Sites that require user credentials (username and password) for access but do not implement advanced anti-bot measures.
\item \textbf{Complex Authentication}: Sites that enforce stronger verification mechanisms, such as multi-factor authentication. In our experiment, {\em email verification} was used as a second authentication layer.
\item \textbf{CAPTCHA}: Pages from demonstration or testing environments where CAPTCHA challenges are triggered consistently on every access attempt, designed to showcase CAPTCHA workflows.
We used demo CAPTCHA websites to ensure consistent and reproducible CAPTCHAs.
Public websites tend to trigger CAPTCHA intermittently or offer tests of varying complexity when a user exhibits suspicious behavior that is hard to reproduce. 

\end{enumerate}

\subsection{Scraping Workflows}

    \para{LLM-assisted Scripting (LAS).} The user executes code and manages control flow; the LLM assists via code generation. We select BeautifulSoup and Scrapy because they are widely used, well-documented baselines, that are also easy to setup and run. We don't include browser automation frameworks (e.g., Selenium) because they require extensive setup which can be challenging for beginners.
    
    \vspace{4pt}
    \begin{itemize}[leftmargin=20pt, itemsep=4pt, label={$\triangleright$}]
        
        \item \textbf{BeautifulSoup.} 
        Lightweight parser for largely static markup; user handles HTTP fetching; no client-side execution; extraction via tags or selectors.
        
        \item \textbf{Scrapy.} 
        Full crawling framework with link following, sessions/cookies, and middleware; logins implemented in spiders; default fetching without running client-side scripts.
        
    \end{itemize}
    
    \vspace{8pt}
    
    \para{End-to-End LLM Agent (ELA).} The agent plans and acts using internal tools; the user supplies a goal prompt and typically intervenes only on failure. We select general-purpose agent Claude (tool-enabled agent), and Simular.ai, which is an agent specialized for web crawling.
    
    \vspace{4pt}
    \begin{itemize}[leftmargin=20pt, itemsep=4pt, label={$\triangleright$}]
        
        \item \textbf{Claude (tool-enabled agent).} 
        A general LLM acting as an agent with built-in tools (e.g., browsing/DOM access, structured extraction, retries/error checks) to complete end-to-end tasks from a goal prompt.
        
        \item \textbf{Simular.ai.} 
        A visual web agent that operates a real browser; performs element selection, clicking, typing, and navigation with integrated retries and error detection; suited for dynamic, script-heavy sites and interactive flows.
        
    \end{itemize}

\subsection{Standardization and Control}
\para{Execution Environment.} All tools ran on the same MacOS workstation under consistent network conditions. BeautifulSoup 4.13.4 and Scrapy 2.13.1 used Python 3.10 with site-specific scripts. Claude 3.5 Sonnet and Simular.ai 0.10.16 were accessed via their websites with natural language prompts. Each tool extracted predefined data elements (e.g., article titles, product prices, and user profile names). For login-protected sites, credentials were hardcoded where needed. We recorded whether the tool (1) accessed the target page, (2) extracted the intended content, and (3) saved results in a usable CSV format. 
To eliminate confounding variables, each scraping task was defined in advance and held constant across all tools. For example, when testing login functionality, the task would be to authenticate, reach the dashboard, and extract a visible headline.
Each site-tool pair was tested up to three times to account for variability in internet response, site behavior, or LLM performance. Inconsistent behavior (e.g., partial loading, rate limiting) was noted and categorized accordingly. Finally, all tools were used in non-incognito, default browser sessions with no browser automation or third-party extensions. Sessions were cleared between runs to avoid caching or cookie reuse bias. Screenshots, error logs, and resulting data files were saved for each attempt. We mention all the prompts we used in the appendix.

\section{Evaluation and Results}
\label{sec:evaluation}

In this section, we evaluate the performance of traditional web scraping tools and modern LLM/computer-use agents using our proposed benchmarking framework of 35 websites.

\subsection{Setup}

\begin{table*}[ht]
\centering
\small
\begin{tabular}{lcccc}
\toprule
Category & BeautifulSoup & Scrapy & Claude & Simular.ai \\
\midrule
Simple HTML & 0.93 & 0.82 & 1.00 & \textbf{1.00} \\
Complex HTML & 0.80 & 0.20 & 0.57 & \textbf{1.00} \\
Simple Authentication & Not Supported & Not Supported & 0.20 & \textbf{0.63} \\
Complex Authentication & Not Supported & Not Supported & 0.12 & \textbf{0.70} \\
CAPTCHA & Not Supported & Not Supported & 0.05 & \textbf{0.10} \\
\bottomrule
\end{tabular}
\caption{Extraction Success Rate (ESR) across benchmark categories (average of 3 trials). ``Not Supported'' means the task could not be completed with the basic prompts used in our tests (mimicking lay users' prompting capabilities). 
}
\label{tab:overall-esr}
\end{table*}

\para{Evaluation Metrics.} We used the following metrics:
\begin{itemize}
    \item \textbf{Extraction Success Rate (ESR):} \% of scraping runs with complete and correct output.
    \item \textbf{Execution Time:} Average runtime for a successful scrape.
\item \textbf{Manual Effort Required (MER).} Quantifies human intervention needed to reach a successful run. We track the (i) Retry Count and (ii) Effort in setting up. Lower manual effort indicates better out-of-the-box reliability and faster adoption.

\textbf{MER Scale:} \\
\textbf{High} --- 5+ retries, substantial manual debugging, or extensive tool-specific setup.\\
\textbf{Medium} --- 2–4 retries, prompt tuning, or routine authentication/session handling.\\
\textbf{Low} --- 0–1 retries; effectively run-and-go with no manual edits. 
\end{itemize}

\begin{table*}[ht]
\centering
\small
\begin{tabular}{lcccc}
\toprule
Category & BeautifulSoup & Scrapy & Claude & Simular.ai \\
\midrule
Simple HTML & $\bullet\bullet$ & $\bullet\bullet$ & $\bullet$ & $\bullet$ \\
Complex HTML & $\bullet\bullet$ & $\bullet\bullet$ & $\bullet\bullet$ & $\bullet$ \\
Simple Authentication  & Not Supported & Not Supported & $\bullet\bullet\bullet$ & $\bullet\bullet$ \\
Complex Authentication  & Not Supported & Not Supported & $\bullet\bullet\bullet$ & $\bullet\bullet$ \\
CAPTCHA & Not Supported & Not Supported & $\bullet\bullet\bullet$ & $\bullet\bullet\bullet$ \\
\bottomrule
\end{tabular}
\caption{Manual Effort Required ($\bullet$=Low, $\bullet\bullet$=Medium, $\bullet\bullet\bullet$=High effort based on average retries, prompts, or debugging). 
"Not Supported" indicates that the framework cannot be used to reasonably automate the scenario without extensive effort or expert guidance.
}
\label{tab:effort-table}
\end{table*}

\para{Workflow.} We used a structured workflow reflecting real scraping tasks. For BeautifulSoup and Scrapy, prompts (manual or notebook-assisted) generated initial code, which was iteratively reviewed, executed, and revised for selectors, sessions, and logins. Once outputs were validated with minimal tuning, the final scripts were tested on benchmark sites. Simular.ai and Claude were evaluated via direct interaction: GUI instructions (Simular.ai) or natural language commands (Claude), with no code modification. For all tools, we logged each attempt, recorded success, execution time, and manual effort, and documented prompt–response patterns. This comparison highlights the contrast between code-driven and agent-driven workflows.

\para{Controlled Environment.}
All experiments were run on the same MacOS workstation with 8-core CPU, 16GB RAM, and stable Wi-Fi. Each scraping method was executed three times per site with randomized delays to simulate human-like behavior. All outputs were saved in CSV for comparison. For consistency, scraping runs were conducted within the same 72-hour window, using identical versioned codebases and avoiding time-based bias.
Each tool was provided with equivalent input tasks, such as logging in, reaching a target page, and extracting content. Minimal tuning was allowed to avoid overfitting.

\subsection{Results}
\label{sec:results}

\para{Agents excel on complex and protected websites.} Table~\ref{tab:overall-esr} summarizes extraction success rates across all five difficulty categories. End-to-end agents dramatically outperformed LLM-assisted scripting on sites with authentication, protection mechanisms, and CAPTCHA challenges. Simular.ai achieved the highest overall performance, with perfect success rates (1.00) on both simple and complex HTML, 63\% success on simple authentication sites, and 70\% on complex authentication pages, and low success on CAPTCHA pages. Claude comparatively underperformed on authentication workflows (20\% on Simple) and achieved 5\% success on dynamic CAPTCHA sites, while maintaining perfect scores on simple HTML.

\begin{figure}[t]
    \centering
\includegraphics[width=\columnwidth]{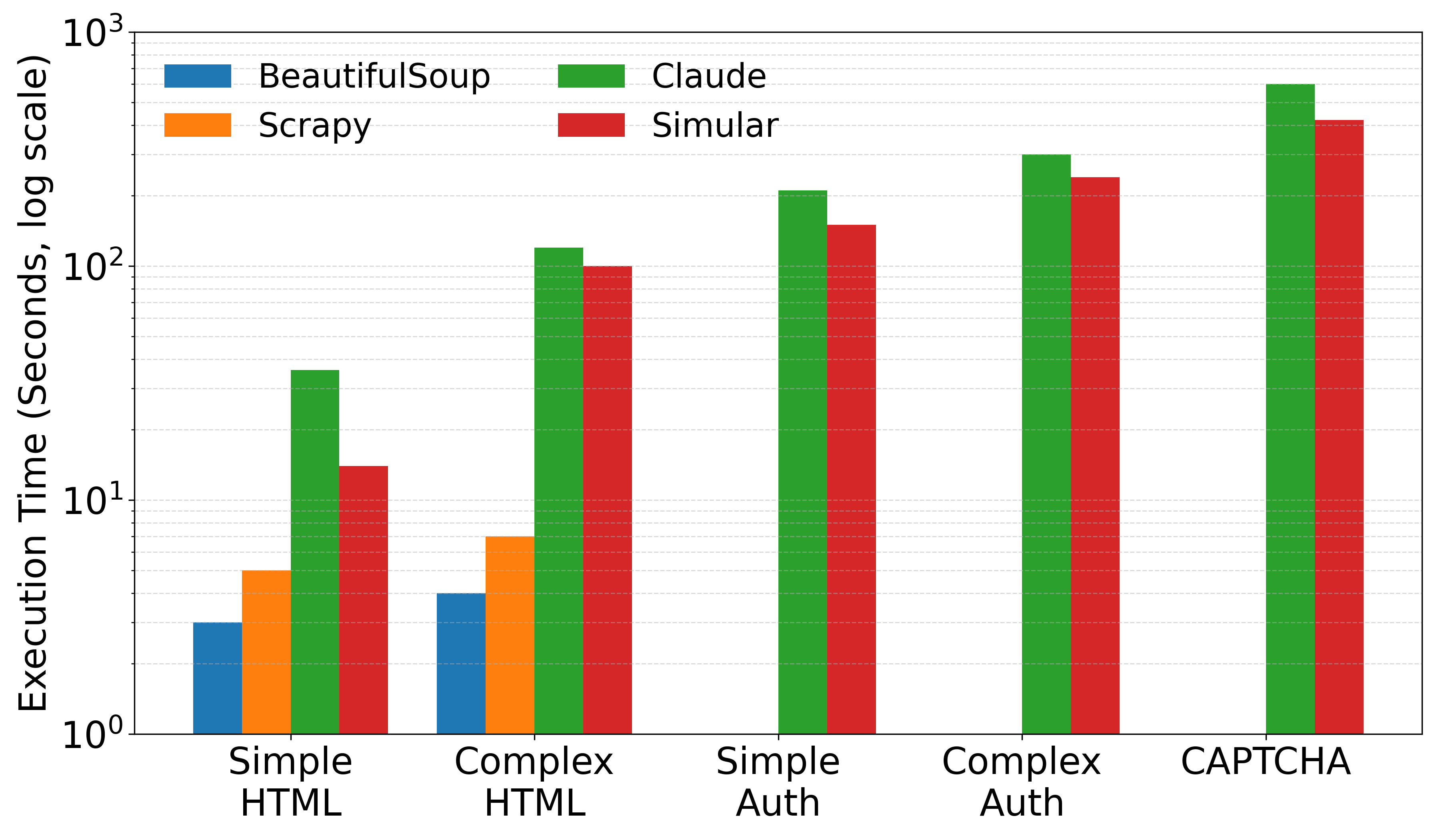}
    \caption{Average execution time per category. Note that the y-axis is on a log scale (in seconds). } 
\label{fig:mt_bench_incrase}
\end{figure}

\para{Traditional approaches remain superior for static content.} Figure~\ref{fig:mt_bench_incrase} reveals execution times that demonstrate a clear trade-off. On simple HTML sites, LLM-assisted scripting achieved high success rates (93\% for BeautifulSoup, 82\% for Scrapy) while completing tasks in under 2 seconds. End-to-end agents matched or exceeded these success rates (100\% for both Claude and Simular.ai) but required 10--20 seconds due to browser automation overhead. For static content, the 15--20× runtime cost of agents provides no meaningful benefit.

\para{Manual effort varies dramatically by workflow complexity.} Table \ref{tab:effort-table} shows how much work users need to put in with each tool. For simple HTML sites, BeautifulSoup and Scrapy needed medium effort (2-4 tries with some tweaking), while Claude and Simular.ai worked with low effort (0-1 tries, basically plug-and-play). On complex HTML sites, traditional tools still needed medium effort, while agents stayed at low effort. The big difference shows up on sites with logins and CAPTCHAs: traditional tools simply don't work at all, while agents need high effort (Claude required 5+ tries with lots of debugging, Simular.ai needed 2-4 tries with standard setup). This shows that while agents might be slower, they actually work on complex sites where traditional tools completely fail, making them much more accessible for beginners who can't handle extensive debugging and technical setup.

\section{Discussion}
\label{sec:discussion}

For static HTML sites, using traditional LLM-assisted scripting tools such as BeautifulSoup or Scrapy can deliver high success rates with minimal overhead. Meanwhile, end-to-end agents can be applied to more complex scenarios such as sites with authentication, heavy JavaScript, or CAPTCHA. For those sites, while traditional tools fall short, LLM agents can still offer viable, if slower, solutions. In addition, future work can explore how to combine the strengths of traditional and LLM-assisted scraping tools to maximize their effectiveness, efficiency, and usability. For example, LLM-assisted tools can be used to handle the initial authentication step, and then traditional tools can take the obtained session keys for efficient content scraping. 
From the defenders' perspective, to develop effective anti-scraping mechanisms, researchers need to proactively consider the capabilities of LLM agents to counter their weaknesses.


\section{Conclusion and Future Work}

In this paper, we evaluated two LLM-powered frameworks used by everyday users across 35 websites spanning five increasingly complex categories of webpages. We find a clear division of strengths between LLM-assisted scripting and end-to-end agents. Traditional scripting frameworks, when paired with LLMs, remain the fastest and most efficient choice for static or lightly dynamic sites, but they completely break down once authentication or protection mechanisms are introduced. End-to-end agents, while slower and often requiring more retries, are the only approaches that consistently succeed on complex and protected websites. Overall, this highlights a trade-off: scripting is superior for static extraction, but agents are useful for navigating dynamic content, authentication workflows, and CAPTCHAs.

This study is limited by a single hardware environment, fixed tool versions, and a small set of testing websites. Future work may consider enlarging the corpus to cover more diverse websites, especially those with stronger anti-bot defenses. Another future direction is to explore a hybrid approach where LLM agents are used to bypass the authentication measures, while LLM-assisted Scripting with traditional methods is used for the actual scraping. Additionally, further research can be done to explore context-aware scripting for web scraping, to understand the specific context and structure for each website, to make it more robust for diverse scenarios. Finally, researchers may study stronger anti-bot solutions given the new capabilities of LLM-powered web scripting.

\section{Acknowledgment} 
This work was in part supported by the Frontier Model Forum (FMF 122987) and the National Science Foundation (NSF) under grant 2229876.

\bibliography{AAAI2026_Final}

\appendix
\section{Appendix}\label{sec:appendix}

\begin{lstlisting}[
    boxpos=t,
    caption={\textbf{Claude and Simular prompt for simple HTML/Complex HTML benchmarking}}, 
    label={lst:claude_simular_simple_html}
]

Load the home pages of the following websites, one by one:
[
  https://thehackernews.com/,
  https://www.wikipedia.org/,
  https://www.investopedia.com/,
  https://www.goodreads.com/
]
Note : For both Simple HTML and Complex HTML prompt remains same, only list of website changes as mentioned in table. We gave exactly same prompts for Claude and Simular.ai

For each website:
1. Record the start time.
2. Navigate to the home page.
3. Confirm that the page has loaded successfully by checking for a known keyword (e.g., \"Hacker News\", \"Wikipedia\") or a visible page title.
4. Record whether the attempt is a Success or Fail.
5. Record the total time taken in seconds.

After testing all websites, create a CSV file named \"Simple_results.csv\"
with the following columns:

Website, SuccessFail, TimeTakenSeconds, ValidationCriteriaUsed

If any website fails to load within 20 seconds, mark it as Fail and continue to the next website. Do not stop the overall process.
\end{lstlisting}

\begin{lstlisting}[
    boxpos=t,
    caption={\textbf{Claude/Simular.ai prompt for simple Authentication/Complex Authentication benchmarking}}, 
    label={lst:claude_simular_Authentication}
]
Load the login pages of the following websites, one by one. For each website login provide username as  <> and Password as <>
[
  https://www.paypal.com/,
  https://www.ebay.org/,
  https://www.youtube.com/,
  https://www.netflix.com/,
  ......,
]
Note : For both simple authentication and complex authentication prompt remains the same, only list of website changes as mentioned in Table.For all these websites we created test accounts with same username and passwords.We gave exactly same prompts for Claude and Simular
For each website:
1. Record the start time.
2. Validate that the homepage loaded successfully by checking for your profile name and header of home page after login
3. Record whether the attempt is a Success or Fail.
4. Record the total time taken in seconds.

After testing all websites, create a CSV file named \"Authentication_results.csv\"
with the following columns:

Website, SuccessFail, TimeTakenSeconds, ValidationCriteriaUsed

If any website fails to load within 50 seconds, mark it as Fail and continue to the next website. Do not stop the overall process.
\end{lstlisting}

\begin{lstlisting}[
    boxpos=t,
    caption={\textbf{Claude/Simular.ai prompt for CAPTCHA}}, 
    label={lst:claude_simular_capcha}
]
1. Navigate to https://www.google.com/recaptcha/api2/demo 
2.Solve the recaptcha challenge make sure you select "I am not robot".
3.Click the submit button
4. Confirm the success message appears and page is loaded
5.create a CSV file named \"Capcha_results.csv\"
with the following columns:
Website, SuccessFail, TimeTakenSeconds, ValidationCriteriaUsed
If any website fails to load within 15 minutes, mark it as Fail and continue to the next website. Do not stop the overall process.
\end{lstlisting}

\begin{lstlisting}[
    boxpos=t,
    caption={\textbf{Sample prompt for Complex/MFA using email for Authentication }}, 
    label={lst:MFA_Authentication}
]
1. Navigate to https://www.expedia.com login page.
2. Enter credentials:
   - Email: <YOUR_TEST_EMAIL>
   - Password: <YOUR_TEST_PASSWORD>
3. Wait for the 6-digit MFA code prompt.
4. Switch to Gmail:
   - Log in with <YOUR_TEST_EMAIL> and <YOUR_TEST_PASSWORD>.
   - Retrieve the latest Expedia email containing the 6-digit code.
5. Return to Expedia, enter the code, and complete login.

Note : We gave exactly same prompts for Claude and Simular
For each website:
1. Record the start time.
2. Validate that the expedia homepage loaded successfully by checking for your profile name and header of home page afterlogin
3. Record whether the attempt is a Success or Fail.
4. Record the total time taken in seconds.
After testing all websites, create a CSV file named \"MFA_results.csv\"
with the following columns:
Website, SuccessFail, TimeTakenSeconds, ValidationCriteriaUsed
If any website fails to load within 500 seconds, mark it as Fail and continue to the next website. Do not stop the overall process.
\end{lstlisting}

\end{document}